%% file: LinearFI-current.tex
\documentclass[twocolumn,aps,prd,superscriptaddress,nofootinbib,showpacs]{revtex4-1}
\usepackage{graphicx}
\usepackage{epstopdf}
\usepackage{color}
\graphicspath{{figs/}{fig/}}
\usepackage{amsmath}
\usepackage{bm}
\usepackage{slashed}
\usepackage{epsfig}
\usepackage{amsfonts}
\usepackage{epstopdf}
\usepackage{hyperref}
\usepackage{float}

\newcommand{\sect}[1]{\section{#1}}
\newcommand{\mi}{\mathrm{i}}
\newcommand{\md}{\mathrm{d}}
\newcommand{\vwJ}{\overrightarrow{{\hspace{-1mm}\widetilde J}}\hspace{-1mm}}
\newcommand{\vhJ}{\overrightarrow{{\hspace{-1mm}\widehat J}}\hspace{-1mm}}

\begin{document}
\title{Automatic computation of Feynman integrals containing linear propagators via auxiliary mass flow}

\author{Zhi-Feng Liu}
\email{xiangshui@pku.edu.cn}
\affiliation{School of Physics and State Key Laboratory of Nuclear Physics and
Technology, Peking University, Beijing 100871, China}
\author{Yan-Qing Ma}
\email{yqma@pku.edu.cn}
\affiliation{School of Physics and State Key Laboratory of Nuclear Physics and
Technology, Peking University, Beijing 100871, China}
\affiliation{Center for High Energy Physics, Peking University, Beijing 100871, China}
\affiliation{Collaborative Innovation Center of Quantum Matter,
Beijing 100871, China}

\begin{abstract}
We proposed a recipe to systematically calculate Feynman integrals containing linear propagators using the auxiliary mass flow method. The key of the recipe is to introduce a quadratic term for each linear propagator and then using differential equations to get rid of their  effects. As an application, we calculated all master integrals of vacuum integrals containing a gauge link up to four loops, and we checked the results by nontrivial dimensional recurrence relations.
\end{abstract}

\maketitle
\allowdisplaybreaks

\sect{Introduction}
As particle physics enters the era of precision, it is necessary to compute perturbative quantum field theory to high order in coupling constants. To this end, the evaluation of Feynman integrals is an unavoidable task. The main strategy at present is to decompose all Feynman integrals in a problem to a small set of bases, called master integrals (MIs), using integration-by-parts (IBP) reduction~\cite{Chetyrkin:1981qh, Laporta:2000dsw, Gluza:2010ws, Schabinger:2011dz, vonManteuffel:2012np, Lee:2013mka, vonManteuffel:2014ixa, Larsen:2015ped, Peraro:2016wsq, Mastrolia:2018uzb, Liu:2018dmc,  Guan:2019bcx, Klappert:2019emp, Peraro:2019svx, Frellesvig:2019kgj, Wang:2019mnn, Smirnov:2019qkx, Klappert:2020nbg, Boehm:2020ijp, Basat:2021xnn, Heller:2021qkz, Bendle:2021ueg}, and then calculate these MIs.

Many methods exist to calculate MIs~\cite{Hepp:1966eg, Roth:1996pd, Binoth:2000ps, Heinrich:2008si, Smirnov:2015mct, Borowka:2015mxa, Borowka:2017idc,Boos:1990rg, Smirnov:1999gc, Tausk:1999vh, Czakon:2005rk, Smirnov:2009up, Gluza:2007rt,Laporta:2000dsw, Lee:2009dh,Kotikov:1990kg, Kotikov:1991pm, Remiddi:1997ny, Gehrmann:1999as, Argeri:2007up, MullerStach:2012mp, Henn:2013pwa, Henn:2014qga, Moriello:2019yhu, Hidding:2020ytt,Catani:2008xa,Rodrigo:2008fp,Bierenbaum:2010cy,Bierenbaum:2012th,Tomboulis:2017rvd,Runkel:2019yrs,Capatti:2019ypt,Aguilera-Verdugo:2020set,Song:2021vru,Dubovyk:2022frj,Liu:2017jxz,Liu:2020kpc,Liu:2021wks,Liu:2022chg}. Among them, the auxiliary mass flow (AMF) method~\cite{Liu:2017jxz,Liu:2020kpc,Liu:2021wks,Liu:2022chg}, a special case of differential equations (DEs) method, calculates MIs by setting up and solving DEs of MIs with respect to an auxiliary mass term $\eta$ (denoted as $\eta$-DEs). This is systematic and efficient, as far as $\eta$-DEs can be set up by using reduction strategy.
Equipped with an iterative strategy~\cite{Liu:2021wks} and the block-triangular reduction form~\cite{ Liu:2018dmc, Guan:2019bcx}, the AMF method becomes so powerful
that many Feynman integrals in cutting-edge problems, which are very challenging for other methods, can be calculated (see Fig. 4 in Ref.~\cite{Liu:2021wks}).
At the same time, the AMF method has already been used in many physical processes \cite{Zhang:2018mlo,Zhang:2020atv,Yang:2020msy,Bronnum-Hansen:2020mzk,Bronnum-Hansen:2021olh,Wu:2021tzo,Bronnum-Hansen:2021pqc,Baranowski:2021gxe}.

The AMF method is systematic only if boundary conditions of the $\eta$-DEs at $\eta \to \infty$ can be systematically obtained. If all inverse propagators defining Feynman integrals are quadratic in loop momenta, where these kinds of Feynman integrals are called {\it quadratic integrals}, then regions at $\eta \to \infty$ can be systematically identified~\cite{Beneke:1997zp, Smirnov:1999bza}, and the contribution of each region can be calculated within the AMF framework~\cite{Liu:2022mfb}. However, life becomes not that easy when some inverse  propagators defining Feynman integrals are linear in loop momenta, where these kinds of Feynman integrals are called {\it linear integrals}. In this case, identification of regions at $\eta \to \infty$ is not that systematic, and the calculation of each region becomes nontrivial~\cite{Zhang:2018mlo,Zhang:2020atv}.
As linear integrals show up frequently in region expansion ~\cite{Beneke:1997zp} and in effective field theories, it will be good if they can be calculated systematically by the powerful AMF method.

In this paper, by introducing an auxiliary quadratic term for each linear propagator, we develop a recipe to calculate any linear integral systematically within the AMF framework.
In the rest of the paper, we first present the general idea of our recipe and then give some examples to illustrate its details. As an application, we calculate
all master integrals of four-loop
vacuum integrals containing a gauge link for the first time, which are useful for studying parton distribution functions~\cite{Ji:2013dva,Ji:2014gla,Radyushkin:2017cyf,Ma:2014jla,Ma:2017pxb}. All results in this paper have been checked by using dimensional recurrence relations~\cite{Tarasov:1996br,Tarasov:2000sf,Lee:2009dh}.


\sect{The  recipe}
Consider a general Feynman integral
\begin{equation}
{I}_{\vec{\nu}}=\prod_{j=1}^L\int\frac{\md^D\ell_j}{\mi\pi^{D/2}}
\prod_{a=1}^{N}\mathcal{D}_a^{-\nu_a},
\end{equation}
where $L$ is the number of loops and $N$ is the number of propagators and irreducible scalar products, and $\mathcal{D}_a$ are inverse propagators
and $\vec{\nu}=(\nu_1,\nu_2,\cdots,\nu_N)$ integers.
We introduce a corresponding auxiliary quadratic integral ${\widetilde I}_{\vec{\nu}}(x)$ defined by
\begin{equation}
{{\widetilde I}}_{\vec{\nu}}(x)=\prod_{j=1}^L\int\frac{\md^D\ell_j}{\mi\pi^{D/2}}
\prod_{a=1}^{N}\widetilde{\mathcal{D}}_a^{-\nu_a}.
\end{equation}
If $\mathcal{D}_a$ is an irreducible scalar product or is quadratic in loop momenta, then $\widetilde{\mathcal{D}}_a={\mathcal{D}}_a$. Otherwise, it can be generally  expressed as
\footnote{If $\mathcal{D}_a$ has a negative infinitesimal imaginary part, we can
simply substitute it by $-\mathcal{D}_a$, leaving an overall factor of $-1$. }
\begin{align}
\mathcal{D}_a &= p_a\cdot\left(\sum_{j=1}^L c_{aj} \ell_j\right)+\Delta_a+\mi 0^+\notag\\
&\equiv  p_a\cdot{\ell}_a+\Delta_a+\mi 0^+,
\end{align}
where $p_a$ is a fixed Lorentz vector, and $\Delta_a$ and $c_{aj}$ are fixed Lorentz scalars. We then define
\begin{align}
\widetilde{\mathcal{D}}_a&=x \left({\ell}_a^{~2}+\mi 0^+\right)+p_a\cdot {\ell}_a+\Delta_a \notag\\
&=x \left[\left({\ell}_a+\frac{p_a}{2x} \right)^2+\frac{\Delta_a}{x}- \frac{p_a^2}{4x^2}+\mi 0^+\right].
\end{align}
Clearly, the original integral can be obtained by
\begin{equation}
{I}_{\vec{\nu}}=\lim_{x\to 0^+}{{\widetilde I}}_{\vec{\nu}}(x),
\end{equation}
which correctly recovers the Feynman prescription.

Denoting MIs of the obtained auxiliary quadratic integrals as $\vwJ(x)$, by using IBP reduction,
we can get the following reduction relation \footnote{With IBP reduction, we in fact only need to calculate MIs of original linear integrals. Then, we can always choose ${{\widetilde I}}_{\vec{\nu}}(x)$ as one of the MIs of auxiliary quadratic integrals.},
\begin{align}\label{eq:red}
{{\widetilde I}}_{\vec{\nu}}(x)=\sum_{i=1}^n c_{\vec{\nu};i}(x) {{\widetilde J}}_i(x),
\end{align}
valid for any value of $x$, where $n$ is the number of MIs and $c_{\vec{\nu};i}(x)$ are rational functions of $x$. Here, ${{\widetilde I}}_{\vec{\nu}}(0^+)$ can be achieved once $\vwJ(x)$ are known. Again using IBP, we can set up a system of DEs for MIs with respect to $x$,
\begin{align}\label{eq:deq}
\frac{\partial}{\partial x}\vwJ(x) = A(x)\vwJ_(x),
\end{align}
where $A(x)$ is a $n\times n$ matrix with elements being rational functions of $x$.
Because Feynman integrals have Feynman parametric representation, then $\vwJ(x)$ have the following asymptotic expansion (see e.g. Ref. \cite{Henn:2014qga})
\begin{equation}\label{eq:miexp}
{{\widetilde J}}_k(x) = \sum_{s} \sum_{i=0}^{n_{s}} x^{s} \ln^{i}(x) \sum_{j=0}^{\infty} {{\widetilde J}}_{k}^{ s \, i \, j} x^j \, ,
\end{equation}
where $s$ is a linear function of space-time dimension $D=4-2\epsilon$ and $n_s$ an integer determined by $s$, and ${{\widetilde J}}_{k}^{ s \, i \,j }$ are functions of $\epsilon$. Taking the limit $x\to 0^+$ is then straight forward \footnote{We assume that dimensional regularization can regularize all singularities of $x$. Therefore,
all terms where $s$ depends on $\epsilon$ can be ignored as $x\to 0^+$. Furthermore, if $s$ is independent of $\epsilon$, it can be only non-negative integers, and $n_s=0$.}
\begin{align}
{{\widetilde J}}_k(0^+) &={{\widetilde J}}_{k}^{ 0 \, 0 \, 0} \, ,\\
{{\widetilde J}}_k^{\prime}(0^+) &={{\widetilde J}}_{k}^{ 0 \, 0 \, 1} \, ,
\end{align}
and so on. Then based on Eq.~\eqref{eq:red}, we can get any desired integral ${I}_{\vec{\nu}}={{\widetilde I}}_{\vec{\nu}}(0^+)$.

Taking advantage of the DEs \eqref{eq:deq}, there are only $n$ independent  ${{\widetilde J}}_{k}^{ s \, i \,j }$, which can be fixed by matching the expansion \eqref{eq:miexp} with values at a given point $\vwJ(x_0)$, as far as $x_0$ is chosen sufficiently small so that the above expansion is convergent at $x=x_0$. Therefore, once $\vwJ(x_0)$ is known, the asymptotic expansion of $\vwJ(x)$ around $x=0$ can be achieved, and thus, ${I}_{\vec{\nu}}$ can be obtained.

For a given regular point $x=x_0$, $\vwJ(x_0)$ are quadratic integrals that can be calculated by using the AMF method \cite{Liu:2017jxz,Liu:2020kpc,Liu:2021wks,Liu:2022chg}. A convenient choice can be replacing $\mi 0^+$ by $-\eta+\mi 0^+$ in all modified propagators, i.e., propagators which are linear in loop momenta when $x\to 0^+$. The advantage of this choice is that the number of MIs does not increase after introducing $\eta$. As $\eta \to \infty$, the newly defined integrals in the AMF method are essentially single-mass vacuum integrals~\cite{Liu:2021wks},
which can be easily calculated within the AMF framework \cite{Liu:2022mfb}.
With this boundary condition, we can obtain $\vwJ(x_0)$ by flowing $\eta$ from $\infty$ to $\mi0^-$ using $\eta$-DEs.

As numerically solving DEs is very efficient, our method is thus systematic and efficient as far as DEs can be set up.

\begin{figure}[H]
	\centerline{\includegraphics[width=0.25\textwidth]{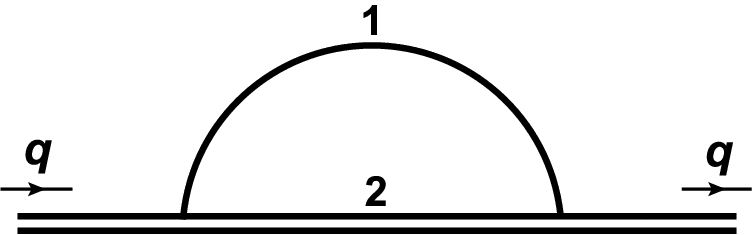}}
	\caption{One-loop vacuum integral containing a gauge link. Double line denotes the gauge link.}\label{fig:ga1-new-tag}
\end{figure}

\sect{A pedagogical example}
To elaborate how to use the above proposed recipe to calculate linear integrals, we take the
simplest linear integral with one linear propagator and
one quadratic propagator as an example,
\begin{align}
{I}_{(1,1)}=\int\frac{\md^D\ell}{\mi\pi^{D/2}}\frac{1}{(\ell^2+\mi0^+)(\ell\cdot n+q\cdot n+\mi0^+)},
\end{align}
where $\ell$ is the loop momentum, and $q$ and $n$ are external momenta with $n^2=-1$.  Clearly, this integral only depends on the scale $q\cdot n$. Without loss of generality, we choose $q\cdot n=1$. This Feynman integral is relevant to the Feynman diagram shown in Fig. \ref{fig:ga1-new-tag}, where the double line stands for a gauge link which contributes linear propagators. We call this a one-loop bubble integral because there is no external legs of particles.

The corresponding auxiliary quadratic integral is defined as
\begin{align}
{\widetilde I}_{(1,1)}(x)=\int\frac{\md^D\ell}{\mi\pi^{D/2}}\frac{1}{\widetilde{\mathcal{D}}_1 \widetilde{\mathcal{D}}_2},
\end{align}
with
\begin{align}
\widetilde{\mathcal{D}}_1=&\ell^2+\mi0^+,\\
\widetilde{\mathcal{D}}_2=&x\left[ \left(\ell+\frac{n}{2x}\right)^2+\frac{1+4x}{4x^2}+\mi0^+\right].
\end{align}
For finite $x$, there are two MIs which can be defined as
\begin{align}
{\widetilde J}_1(x)&=\int\frac{\md^D\ell}{\mi\pi^{D/2}}\frac{x}{\widetilde{\mathcal{D}}_2},\\
{\widetilde J}_2(x)&=\int\frac{\md^D\ell}{\mi\pi^{D/2}}\frac{x}{\widetilde{\mathcal{D}}_1 \widetilde{\mathcal{D}}_2},
\end{align}
with ${\widetilde I}_{(1,1)}(x)=\frac{{\widetilde J}_2(x)}{x}$.

Using IBP, we get the system of $x$-DEs,
\begin{equation}  \label{eq:des1}
\frac{\partial}{\partial x}\vwJ(x)=\left(
\begin{array}{cc}
\frac{2 (-1 + \epsilon) (1 + 2 x)}{x (1 + 4 x)} & 0 \\  %
-\frac{4 (-1 + \epsilon) x}{1 + 4 x} & \frac{1}{x}
\end{array}
\right)\vwJ(x).
\end{equation}
Besides $x=\infty$, there are two  additional singularities $x=0$ and $x=-\frac{1}{4}$, which correspond to an infinity mass term and zero mass term in $\widetilde{\mathcal{D}}_2$, respectively. Therefore, we can choose boundary conditions at, e.g., $x_0=\frac{1}{16}$, which point can be well estimated by asymptotic expansions around $x=0$.
Based on DEs \eqref{eq:des1}, we get the following asymptotic expansions:
\begin{align}
{\widetilde J}_1(x) &=  x^{-2 + 2 \epsilon}\sum_{i=0}^{\infty}a_{i}x^{i} ,\\
{\widetilde J}_2(x) &=  c_{0}x+x^{2 \epsilon}\sum_{i=0}^{\infty}b_{i}x^{i} ,
\end{align}
which gives
\begin{align}
{\widetilde I}_{(1,1)}(x) &=\frac{{\widetilde J}_2(x)}{x}=  c_{0}+x^{2 \epsilon-1}\sum_{i=0}^{\infty}b_{i}x^{i} .
\end{align}
As $x\to0^+$, we find ${\widetilde J}_1(0^+)=0$ and ${\widetilde I}_{(1,1)}(0^+) =c_0$.
The DEs \eqref{eq:des1} also provide recurrence relations of coefficients,
\begin{align}
a_{i}&=\frac{8 - 4 \epsilon - 4 i}{i}a_{i-1},\\
b_{i}&=-\frac{4 (-1 + \epsilon)}{-1 + 2 \epsilon + i}a_{i}-\frac{4 (-2 + 2 \epsilon + i)}{-1 + 2 \epsilon + i}b_{i-1},\\
b_{0}&=-\frac{4 (-1 + \epsilon)}{-1 + 2 \epsilon }a_{0},
\end{align}
where $i\geq 1$. Therefore, only $a_{0}$ and $c_{0}$ are free coefficients.

To fix $a_{0}$ and $c_{0}$,
we need to know $\vwJ(\frac{1}{16})$. Using the package \texttt{AMFlow} \cite{Liu:2022chg}, we get
\begin{align}
&{{\widetilde J}_1}(\frac{1}{16})= - 80.00000000000000\epsilon^{-1}\nonumber\\
&   + (316.7393839660332 -251.3274122871835 \mi)\nonumber\\
&   - (338.0371701002434 - 995.0661217702465 \mi) \epsilon\nonumber\\
&   - (311.267338694263 +1888.809135035203 \mi) \epsilon^2\nonumber\\
&   + (1510.395112926408 + 2295.761140388504 \mi) \epsilon^3\nonumber\\
&   - (2656.967479760546 + 2012.921788373378 \mi) \epsilon^4\nonumber\\
&   + (3337.360161271413 +1359.608268441597 \mi) \epsilon^5\nonumber\\
&   - (3544.782677804504 + 737.597502301953 \mi) \epsilon^6\nonumber\\
&   + (3503.444773310217 +330.827170478017 \mi) \epsilon^7\nonumber\\
&   - (3393.220492080171 +125.304259848717 \mi) \epsilon^8,\\
&{{\widetilde J}_2}(\frac{1}{16})= 1.000000000000000\epsilon^{-1}\nonumber\\
& - (3.361601777683940 -3.141592653589793 \mi)\nonumber\\
& + (2.540630265434342 - 10.560783449066254 \mi) \epsilon\nonumber\\
& + (4.39414268443412 +  18.31705093747636 \mi) \epsilon^2\nonumber\\
& - (15.75459001722795 +20.93897855970841 \mi) \epsilon^3\nonumber\\
& + (25.55137360766245 +17.56661536372871 \mi) \epsilon^4\nonumber\\
& - (31.12100936594129 + 11.47483328701037 \mi) \epsilon^5\nonumber\\
& + (32.43597076695672 + 6.06440829051716 \mi) \epsilon^6\nonumber\\
& - (32.01028557435537 +2.66344640964930 \mi) \epsilon^7\nonumber\\
& + (30.90646894444592 +0.99159094290134 \mi) \epsilon^8,
\end{align}
where we only keep 10 orders in $\epsilon$ expansion with 16-digit precision, although higher order and more digits can be obtained easily. We keep the same accuracy in all examples in this paper.
Calculation of $\vwJ(\frac{1}{16})$ in this example is certainly a very simple problem, but for general complicated problems we can still use the AMF method to calculate them.
By matching the asymptotic expansions  of $\vwJ(x)$ at $x=0$ with $\vwJ(\frac{1}{16})$, we can determine $a_0$ and $c_0$.
Eventually we obtain
\begin{align}
&{I}_{(1,1)}=c_0= - 2.000000000000000\epsilon^{-1}\nonumber\\
& - (0.072979947957153 +   6.283185307179586 \mi)\nonumber\\
& - (2.356397451352731 + 0.229273268361557 \mi) \epsilon\nonumber\\
& + (0.19031305777140 -28.07369204230733 \mi) \epsilon^2\nonumber\\
& - (2.342969999912346 +0.156392715314289 \mi) \epsilon^3\nonumber\\
& + (0.29757525976318 - 113.32027706384633 \mi) \epsilon^4\nonumber\\
& - (2.334468092829592 +0.075946730925850 \mi) \epsilon^5\nonumber\\
& + (0.3248443395722 -   453.6966444901993 \mi) \epsilon^6\nonumber\\
& - (2.333380642220398 + 0.034360215639063 \mi) \epsilon^7\nonumber\\
& + (0.331273935869 -1814.969761821040 \mi) \epsilon^8.
\end{align}

\sect{Vacuum integrals containing a gauge link up to four loops}
Now, we apply our recipe to calculate vacuum integrals containing a gauge link up to four loops.
These results are useful for studying parton distribution functions~\cite{Ji:2013dva,Ji:2014gla,Radyushkin:2017cyf,Ma:2014jla,Ma:2017pxb}. On the one hand they can serve as boundary conditions for four-loop parton correlation functions calculation,
and on the other hand, they are needed to renormalize three-loop parton correlation functions~\cite{Braun:2018brg,Li:2020xml}.
In general, for an $L$-loop vacuum integral with a gauge link, there must be at least $L$ quadratic propagators
and at most $L$ linear propagators, or else, it is either reducible or scaleless.
With this restriction, up to four loops we find that there are 64 nonzero and irreducible sectors in total which give 65 MIs.
The 64 nonzero and irreducible sectors are displayed in the the supplementary material together with results of all the 65 MIs.

\begin{figure}[H]
	\centerline{\includegraphics[width=0.2\textwidth]{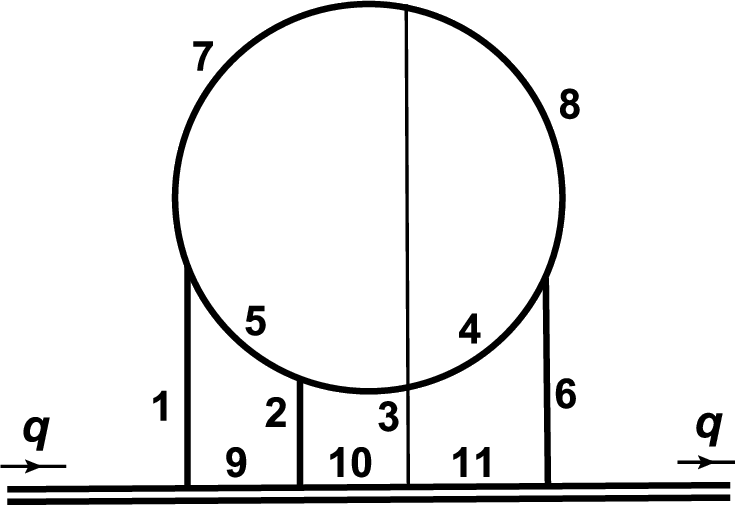}}
	\caption{The most complex integral family for 4-loop vacuum integrals containing a gauge link. Double lines denote the gauge link.}\label{fig:gd41-new-tag}
\end{figure}

We take the most complicated integral family shown in Fig.~\ref{fig:gd41-new-tag} as an example to explain our calculation details. All other families can be calculated similarly and final results are provided in the supplementary material.

A general integral in this family can be expressed as
\begin{equation}
{I}_{\vec{\nu}}=\left(\prod_{j=1}^{4}\int\frac{\md^D\ell_j}{\mi\pi^{D/2}}\right)
\frac{\mathcal{D}_{12}^{-\nu_{12}}\mathcal{D}_{13}^{-\nu_{13}}\mathcal{D}_{14}^{-\nu_{14}}}{\mathcal{D}_{1}^{\nu_{1}}\mathcal{D}_{2}^{\nu_{2}}\cdots \mathcal{D}_{11}^{\nu_{11}}},
\end{equation}
where ${\mathcal{D}_{1},\mathcal{D}_{2},\cdots,\mathcal{D}_{8}}$ are quadratic denominators, ${\mathcal{D}_{9},\mathcal{D}_{10},\mathcal{D}_{11}}$
linear denominators, and ${\mathcal{D}_{12},\mathcal{D}_{13},\mathcal{D}_{14}}$ irreducible numerators, with explicit form
\begin{align}
&\mathcal{D}_1=\ell_1^2,\,\mathcal{D}_2=\ell_2^2,\,\mathcal{D}_3=\ell_3^2,\,\mathcal{D}_4=\ell_4^2,\,\mathcal{D}_5=(\ell_2+\ell_4)^2,\nonumber\\
&\mathcal{D}_6=(\ell_1+\ell_2+\ell_3)^2,\,\mathcal{D}_7=(\ell_1+\ell_2+\ell_4)^2,\nonumber\\
&\mathcal{D}_8=(\ell_1+\ell_2+\ell_3+\ell_4)^2,\,\mathcal{D}_9=(\ell_{1}+q)\cdot n,\nonumber\\
&\mathcal{D}_{10}=(\ell_{1}+\ell_{2}+q)\cdot n,\,\mathcal{D}_{11}=(\ell_{1}+\ell_{2}+\ell_{3}+q)\cdot n,\nonumber\\
&\mathcal{D}_{12}=(\ell_1+\ell_4)^2,\,\mathcal{D}_{13}=(\ell_2+\ell_3)^2,\,\mathcal{D}_{14}=\ell_{4} \cdot n ,\notag
\end{align}
where Feynman prescription  $+\mi0^+$ is omitted. This family has 32 MIs in total. Similar to the one-loop case, the single scale $q\cdot n$ can be extracted as an overall factor, and thus we can simply set $q\cdot n=1$.

Based on our recipe, we define a corresponding auxiliary family of integrals,
\begin{equation}
	{\widetilde I}_{\vec{\nu}}(x)=\left(\prod_{j=1}^{4}\int\frac{\md^D\ell_j}{\mi\pi^{D/2}}\right)
	\frac{\widetilde{\mathcal{D}}_{12}^{-\nu_{12}}\widetilde{\mathcal{D}}_{13}^{-\nu_{13}}
		\widetilde{\mathcal{D}}_{14}^{-\nu_{14}}}{\widetilde{\mathcal{D}}_{1}^{\nu_{1}}
		\widetilde{\mathcal{D}}_{2}^{\nu_{2}}\cdots \widetilde{\mathcal{D}}_{11}^{\nu_{11}}},
\end{equation}
where
\begin{align}
\widetilde{\mathcal{D}}_9&={x}\left[{(\ell_{1}+\frac{n}{2x})^2+\frac{1+4x}{4x^2}}+\mi0^+\right],\notag\\
\widetilde{\mathcal{D}}_{10}&={x}\left[{(\ell_{1}+\ell_{2}+\frac{n}{2x})^2+\frac{1+4x}{4x^2}}+\mi0^+\right],\notag\\
\widetilde{\mathcal{D}}_{11}&={x}\left[{(\ell_{1}+\ell_{2}+\ell_{3}+\frac{n}{2x})^2+\frac{1+4x}{4x^2}}+\mi0^+\right],\notag
\end{align}
and $\widetilde{\mathcal{D}}_i={\mathcal{D}}_i$ for $i\neq 9, 10, 11$. After changing linear propagators to quadratic propagators, the number of MIs  in the auxiliary family increases to 120.
We use  $\vwJ(x)$ to denote these MIs. Each ${\widetilde J}_i(x)$ is defined by a specific list of exponents $\vec{\nu}^i$ via the following relation
\begin{align}\label{eq:cancel}
{\widetilde J}_i(x)=x^{\nu_9^i+\nu_{10}^i+\nu_{11}^i}{\widetilde I}_{\vec{\nu}^i}(x),
\end{align}
where,  for convenience, the factor $x^{\nu_9^i+\nu_{10}^i+\nu_{11}^i}$ is introduced to cancel the overall factor $x$ in
$\widetilde{\mathcal{D}}_9,\,\widetilde{\mathcal{D}}_{10},\,\widetilde{\mathcal{D}}_{11}$.

For a regular point, say $x_0=\frac{1}{16}$, we calculate the corresponding MIs $\vwJ(\frac{1}{16})$ using the package \texttt{AMFlow}\cite{Liu:2022chg} with the mass mode at the first step of the iteration. The calculation is automatic and can be decomposed to the following steps.  First, $\mi0^+$ in $\widetilde{\mathcal{D}}_{9/10/11}$ is replaced by $-\eta+\mi 0^+$.
Then, $\eta$-DEs are set up, and boundary conditions at $\eta\to\infty$ are calculated~\cite{Liu:2021wks,Liu:2022mfb}, where \texttt{Kira}\cite{Klappert:2020nbg} is employed to do IBP reduction. Finally, the $\eta$-DEs are solved to realize the flow of $\eta$ from $\infty$ to $\mi0^-$.

In this problem, $x$-DEs of $\vwJ(x)$ can be derived from $\eta$-DEs during calculating $\vwJ(\frac{1}{16})$, details of which are given in the Appendix.
With boundary conditions at $x_0=\frac{1}{16}$, the $x$-DEs are solved (similar to solve $\eta$-DEs) by using  \texttt{AMFlow}\cite{Liu:2022chg} to obtain results
at $x=0^+$,
which gives final results of our desired linear integrals.
Following is the result of the top-sector corner integral,
\begin{align}
&{I}_{(1,\cdots,1,0,0,0)}= -5.184638775716850\epsilon^{-1}\nonumber\\
& - (35.55435225946754 + 65.15209235723534 \mi)\nonumber\\
& +(124.0703029826581 - 446.7891674459475 \mi) \epsilon\nonumber\\
&+ (1306.855351606906 - 1870.355270319554 \mi) \epsilon^{2}\nonumber\\
&+ (8989.362598245622 - 7095.610423728019 \mi)
\epsilon^{3}\nonumber\\
&+ (43885.08629764386 - 21591.89236479742 \mi) \epsilon^{4}\nonumber\\
&+ (200504.4704938221 - 69609.6949015959 \mi)  \epsilon^{5}\nonumber\\
&+ (852097.4905856483 - 196376.9845468376 \mi) \epsilon^{6}\nonumber\\
&+ (3560145.246623535 - 611937.690718618 \mi) \epsilon^{7}\nonumber\\
&+ (14591453.36136039 - 1726015.80403403 \mi) \epsilon^{8}.
\end{align}
Note that higher precision and higher orders in $\epsilon$ can also be achieved easily.

\begin{figure}[H]
	\centerline{\includegraphics[width=0.25\textwidth]{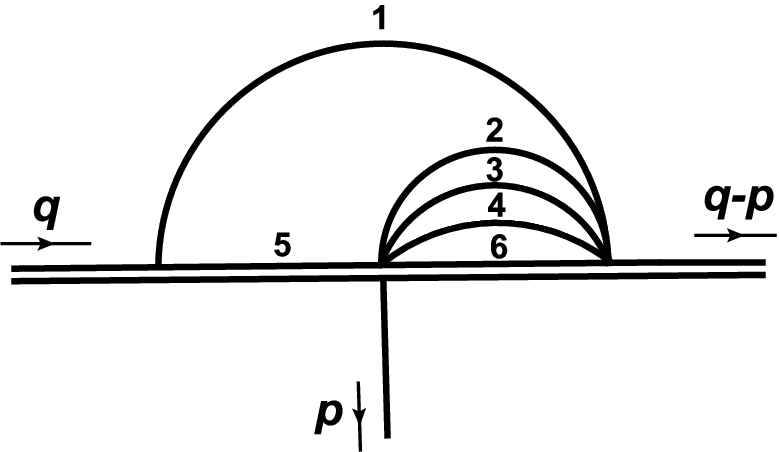}}
	\caption{Four-loop Feynman diagram with a gauge link and an external leg.}\label{fig:gaugelink-legs-new2-tag}
\end{figure}

\sect{Integrals with external legs}
Now let us consider more general linear integrals containing both gauge links and external legs. Take four-loop integrals containing
one external leg shown in Fig. \ref{fig:gaugelink-legs-new2-tag} as an example.
The top-sector corner integral can be expressed as
\begin{equation}
{I}_{(1,1,1,1,1,1)}(q\cdot n,p\cdot n)=
\left(\prod_{j=1}^{4}\int\frac{\md^D\ell_j}{\mi\pi^{D/2}}\right)
\frac{1}{\mathcal{D}_{1}\mathcal{D}_{2}\mathcal{D}_{3}\mathcal{D}_{4}\mathcal{D}_{5}\mathcal{D}_{6}},
\end{equation}
with
\begin{align}
&\mathcal{D}_1=\ell_1^2,\,\mathcal{D}_2=\ell_2^2,\,\mathcal{D}_3=\ell_3^2,\,\mathcal{D}_4=\ell_4^2,\, \mathcal{D}_5=(\ell_1+q)\cdot n,\nonumber\\
&\mathcal{D}_6=(\ell_1+\ell_2+\ell_3+\ell_4+q-p)\cdot n ,
\end{align}
where $p$ is a lightlike external momentum, and Feynman prescription  $+\mi0^+$ for each inverse
propagator is omitted.

To calculate integrals in this example, we define auxiliary quadratic integrals by using  the following inverse propagators
\begin{align}
\widetilde{\mathcal{D}}_5&={x}\left[{(\ell_{1}+\frac{n}{2x})^2+\frac{1}{4x^2}}+\frac{q\cdot n}{x} +\mi 0^+\right],\notag\\
\widetilde{\mathcal{D}}_{6}&={x}\bigg[(\ell_{1}+\ell_{2}+\ell_{3}+\ell_{4}+\frac{n}{2x})^2\notag\\
	&\quad\quad+\frac{1}{4x^2}+\frac{(q-p)\cdot n}{x}+\mi 0^+\bigg],\notag
\end{align}
and $\widetilde{\mathcal{D}}_i={\mathcal{D}}_i$ for  $i\leq 4$.
After changing linear propagators to quadratic propagators, the number of MIs increases from 2 to 6.  Using \texttt{AMFlow} \cite{Liu:2022chg}, we calculate auxiliary quadratic integrals at a fixed point $x=x_0$, and then solve $x$-DEs to obtain integrals at $x=0^+$.
Without loss of generality, for the phase-space point $q\cdot n=1$, $p\cdot n=\frac{1}{2}$,
we obtain
\begin{align}
&{I}_{(1,1,1,1,1,1)}(1,\frac{1}{2})= - 0.001388888888888889\epsilon^{-2}\nonumber\\
& - (0.01673049985543654 + 0.01745329251994330 \mi)\epsilon^{-1}\nonumber\\
& - (0.2181793358910845 + 0.2102416617468981 \mi)\nonumber\\
& - (1.816593593006501 +3.660426889433229 \mi) \epsilon\nonumber\\
& - (15.81900673709547 +33.89466583909418 \mi) \epsilon^2\nonumber\\
& - (110.3704374791621 +401.1363831252734 \mi) \epsilon^3\nonumber\\
& - (810.083299387532 +3287.604829797105 \mi) \epsilon^4\nonumber\\
& - (5155.23653311233 +33468.66058680208 \mi) \epsilon^5\nonumber\\
& - (34800.6654379141 +258370.0763552148 \mi) \epsilon^6\nonumber\\
& - (211074.015357447 +2454130.498362980 \mi) \epsilon^7.
\end{align}
When there are more gauge links or more external legs, the calculation process is similar.

\sect{Self-consistency check by dimensional recurrence relations}
For any integral ${I}_{\vec{\nu}}^D$, where we explicitly write the dependence of spacetime dimension $D$, there is a parametric
representation (see, e.g., \cite{Bitoun:2017nre}),
\begin{align}
{I}_{\vec{\nu}}^D&=\left(\prod_{j=1}^L\int\frac{\md^D\ell_j}{\mi\pi^{D/2}}\right)
\prod_{i=1}^N\mathcal{D}_i^{-\nu_i}\notag\\
&=\left(\prod_{i=1}^N\int_0^{\infty}\frac{x_i^{\nu_i-1}\md x_i}{\Gamma(\nu_i)}
\right)\frac{e^{-F/U}}{U^{D/2}},
\end{align}
and thus,
\begin{equation}
{I}_{\vec{\nu}}^{D-2}=\left(\prod_{i=1}^N\int_0^{\infty}\frac{x_i^{\nu_i-1}\md x_i}{\Gamma(\nu_i)}
\right)\frac{e^{-F/U}}{U^{D/2}}U.
\end{equation}
As $U$ is a homogeneous polynomial of $x_i$, the above expression  can be translated into a linear combination of integrals at dimension $D$ with different $\vec \nu$, which can be reduced to MIs by using IBP identities. Supposing $\vec{J}^D$ is a set of MIs, the above procedure results in dimensional recurrence relations ~\cite{Tarasov:1996br,Tarasov:2000sf},
\begin{equation}\label{eq:drr}
\vec{J}^{D-2}=C^D\vec{J}^D,
\end{equation}
where $C^D$ is a matrix rationally depending on $D$.

Up to three loops, vacuum integrals containing a gauge link have been calculated in Ref.~\cite{Li:2020xml} by solving
dimensional recurrence relations using the method in Ref.~\cite{Lee:2009dh}.
The obtained results agree with ours. For four-loop integrals containing a gauge link, we find it very hard to solve
the corresponding dimensional recurrence relations.
However, the dimensional recurrence relations can provide a highly nontrivial self-consistency check for our results.
We calculate MIs at two different space-time dimensions $D_0$ and $D_0-2$, and we find that the obtained results satisfy
the relations Eq. \eqref{eq:drr} within precision.

\sect{Summary}
In summary, we develop a recipe to calculate linear integrals using the AMF method~\cite{Liu:2017jxz,Liu:2020kpc,Liu:2021wks}. For any given linear integral, our recipe is to introduce an auxiliary quadratic term for each linear propagator, and the obtained auxiliary quadratic integral can be calculated systematically using the AMF method. Taking the result of the auxiliary quadratic integral calculated at fixed auxiliary quadratic terms as the boundary condition and using differential equations to push the auxiliary quadratic terms to zero, effects of  auxiliary quadratic terms  will die out eventually, and we get the result of the target linear integral. This recipe of calculating linear integrals is very systematic and has been implemented in the package~\texttt{AMFlow} \cite{Liu:2022chg}.

As linear integrals show up frequently in region expansion and in effective field theories, our recipe will be useful in phenomenological studies. As the first application, we have calculated all MIs of vacuum integrals containing a gauge link up to four loops,
which are useful to study parton distribution functions. Our results have been checked by nontrivial dimensional recurrence relations.

\sect{Acknowledgments}
We thank K.T. Chao,  X. Liu, Z.Y. Li, X. Li, X. Guan, W.H. Wu, C. Meng, and R.H. Wu for many useful communications
and discussions. The work is supported in part by the National Natural Science Foundation of
China (Grants No. 11875071, No. 11975029), the National Key Research and Development Program of China under
Contract No. 2020YFA0406400, and the High-performance Computing Platform of Peking University.


\appendix

\sect{Derivation of relation between $x$-DEs and $\eta$-DEs for vacuum integrals}
\label{sec:rel}
For  vacuum integrals containing a gauge link, $x$-DEs can be obtained from $\eta$-DEs. We take the family in Fig.~\ref{fig:ga1-new-tag} as an example to explain this, and the final relation is independent of the specific family.
MIs after introducing $x$ are defined by  $\vwJ(x)=\left(\widetilde{J}_1(x), \widetilde{J}_2(x)\right)^T$, with
\begin{align}
&\widetilde{J}_1(x)=\int\frac{\md^D\ell}{\mi\pi^{D/2}}\frac{1}{(\ell+\frac{n}{2x})^2+\frac{1+4x}{4x^2}},\\
&\widetilde{J}_2(x)=\int\frac{\md^D\ell}{\mi\pi^{D/2}}\frac{1}{\ell^2\left((\ell+\frac{n}{2x})^2+\frac{1+4x}{4x^2}\right)},
\end{align}
where Feynman prescription $+\mi0^+$ in each denominator is omitted. The corresponding $x$-DEs are denoted as
\begin{align}
&\frac{\partial}{\partial x}\vwJ (x)=A(x)\vwJ (x).\label{eq:xde}
\end{align}
When calculating $\vwJ (x_0)$ using AMF method, we define modified integrals $\vhJ(\eta)=\left(\widehat{J}_1(\eta), \widehat{J}_2(\eta)\right)^T$, with
\begin{align}
&\widehat{J}_1(\eta)=\int\frac{\md^D\ell}{\mi\pi^{D/2}}\frac{1}{(\ell+\frac{n}{2x_0})^2+\frac{1+4x_0}{4x_0^2}-\eta},\\
&\widehat{J}_2(\eta)=\int\frac{\md^D\ell}{\mi\pi^{D/2}}\frac{1}{\ell^2\left((\ell+\frac{n}{2x_0})^2+\frac{1+4x_0}{4x_0^2}- \eta\right)}.
\end{align}
The corresponding $\eta$-DEs are denoted as
\begin{align}\label{eq:etade}
\frac{\partial}{\partial \eta}\vhJ(\eta)=M(\eta)\vhJ(\eta).
\end{align}
We express $A(x)$ by $M(\eta)$.

Let us define $\overrightarrow{K}(p^2, m^2)=\left(K_1(p^2, m^2), K_2(p^2, m^2)\right)^T$ with
\begin{align}
&K_1(p^2,m^2)=\int\frac{\md^D\ell}{\mi\pi^{D/2}}\frac{1}{(\ell+p)^2+m^2},\\
&K_2(p^2,m^2)=\int\frac{\md^D\ell}{\mi\pi^{D/2}}\frac{1}{\ell^2\left((\ell+p)^2+m^2\right)}.
\end{align}
Because Feynman integrals have a definite mass dimension, there is a rescaling relation,
\begin{align}\label{eq:rescale}
\overrightarrow{K}((1+\delta) p^2, (1+\delta) m^2)=(1+\delta)^\alpha\overrightarrow{K}( p^2,  m^2),
\end{align}
where $\alpha$ is a diagonal matrix with diagonal elements denoting mass dimension. In this example, it is given by
\begin{equation}  \label{eq:des111}
\alpha=\left(
\begin{array}{cc}
\frac{D}{2}-1 & 0 \\
0 & \frac{D}{2}-2
\end{array}
\right).
\end{equation}
Expansion of Eq.~\eqref{eq:rescale} to $O(\delta)$ gives
\begin{align}\label{eq:exp}
\left(p^2 \frac{\partial}{\partial p^2}+m^2 \frac{\partial}{\partial m^2}-\alpha\right)\overrightarrow{K}( p^2,  m^2)=0.
\end{align}

For convenience, we denote
\begin{align}
y&=-\frac{1}{4x^2},\,z=\frac{1+4x}{4x^2},\notag\\
y_0&=-\frac{1}{4x_0^2},\,z_0=\frac{1+4x_0}{4x_0^2}.\notag
\end{align}
Then, integrals are related by
\begin{align}
\vwJ(x)&=\overrightarrow{K}( y,  z),\\
\vhJ(\eta)&=\overrightarrow{K}( y_0,  z_0-\eta).
\end{align}
Based on $\eta$-DEs \eqref{eq:etade}, we get
\begin{align}
\frac{\partial}{\partial \eta} \overrightarrow{K}( y_0,  z_0-\eta)
&=M(\eta)\overrightarrow{K}( y_0, z_0-\eta).
\end{align}
By choosing $\eta=z_0-z$ in the above equation, we get
\begin{align}
\frac{\partial}{\partial z} \overrightarrow{K}( y_0,  z)&=-M(z_0-z)\overrightarrow{K}( y_0,  z),
\end{align}
and thus,
\begin{align}
&\frac{\partial}{\partial z} \overrightarrow{K}( y,  z)\notag\\
=&\frac{\partial}{\partial z}\left[ \left(\frac{y}{y_0}\right)^\alpha \overrightarrow{K}( y_0,  \frac{y_0}{y}z)\right]\notag\\
=&-\left(\frac{y}{y_0}\right)^{\alpha-1} M(z_0-\frac{y_0}{y}z)\overrightarrow{K}( y_0,  \frac{y_0}{y}z)\notag\\
=&-\left(\frac{y}{y_0}\right)^{\alpha-1} M(z_0-\frac{y_0}{y}z) \left(\frac{y}{y_0}\right)^{-\alpha}\overrightarrow{K}( y,  z).
\end{align}
where Eq.~\eqref{eq:rescale} has been used.
Therefore, we have
\begin{align}
&\frac{\partial}{\partial x}\vwJ (x)=\left( \frac{\partial y}{\partial x} \frac{\partial }{\partial y} +\frac{\partial z}{\partial x} \frac{\partial }{\partial z} \right)\overrightarrow{K}( y,  z)\notag\\
&=\left[ \frac{\partial y}{\partial x} \left(\frac{\alpha}{y}-\frac{z}{y} \frac{\partial }{\partial z} \right) +\frac{\partial z}{\partial x} \frac{\partial }{\partial z} \right]\overrightarrow{K}( y,  z)\notag\\
&=\frac{\partial y}{\partial x} \frac{\alpha}{y} \vwJ (x)-\left(\frac{y}{y_0}\right)^{\alpha-1} M(z_0-\frac{y_0}{y}z) \left(\frac{y}{y_0}\right)^{-\alpha}\notag\\
&\hspace{2cm}\times \left[ -\frac{z}{y} \frac{\partial y}{\partial x}  +\frac{\partial z}{\partial x} \right]\vwJ (x),
\end{align}
where the Eq.~\eqref{eq:exp} has been used. The above result means
\begin{align}\label{eq:aeta}
A(x)=&\frac{1}{y}\frac{\partial y}{\partial x} \alpha+\frac{z^2}{y}\left(\frac{\partial}{\partial x}\frac{y}{z}\right)\notag\\ &\times \left(\frac{y}{y_0}\right)^{\alpha-1}M\left(z_0-y_0\frac{z}{y}\right)
\left(\frac{y}{y_0}\right)^{-\alpha},
\end{align}
which connects $x$-DEs to $\eta$-DEs for any vacuum integrals containing a gauge link.
It is useful because $M$ can be more easily obtained.

\input{paper.bbl}
\end{document}

%% file: paper.bbl
\providecommand{\href}[2]{#2}\begingroup\raggedright\endgroup